\documentclass[letterpaper]{article} 
\usepackage{2col}  
\usepackage{times}  
\usepackage{helvet}  
\usepackage{courier}  
\usepackage[hyphens]{url}  
\usepackage{graphicx} 

\usepackage{amssymb}
\usepackage{amsmath} 
\usepackage{multirow}
\newtheorem{theorem}{Theorem}

\newtheorem{condition}[theorem]{Condition}

\newtheorem{corollary}[theorem]{Corollary}

\newtheorem{definition}[theorem]{Definition}

\newtheorem{observation}[theorem]{Observation}

\usepackage{natbib}  
\usepackage{caption} 
\begin{document}

\title{Optimal classification with endogenous behavior}
\author{Elizabeth Maggie Penn\thanks{May 21, 2025. Professor of Political Science and Quantitative Theory \& Methods, Emory University. Email: elizabeth.m.penn@gmail.com. I thank John Patty, Kevin Quinn, and Alexander Tolbert for their comments on this manuscript.} }

\maketitle

\begin{abstract}
 I consider the problem of classifying individual behavior in a simple setting of \textit{outcome performativity} where the behavior the algorithm seeks to classify is itself dependent on the algorithm.  I show in this context that the most accurate classifier is either a \textit{threshold} or a \textit{negative threshold} rule.  A threshold rule offers the ``good" classification to those individuals more likely to have engaged in a desirable behavior, while a negative threshold rule offers the ``good"  outcome to those \textit{less likely} to have engaged in the desirable behavior. While seemingly pathological, I show that a negative threshold rule can maximize classification accuracy when outcomes are performative. I provide an example of such a classifier, and extend the analysis to more general algorithm objectives. A key takeaway is that when behavior is endogenous to classification, optimal classification can negatively correlate with signal information. This may yield negative downstream effects on groups in terms of the aggregate behavior induced by an algorithm.
\end{abstract}

\noindent\textbf{Note (version 2).} \textit{The results in this 2025 preprint have
been generalized and subsumed by E.~M.~Penn and J.~W.~Patty, ``Classification in
Equilibrium: Structure of Optimal Decision Rules'' (2026), arXiv:2612.00375, which
extends them to arbitrary linear designer objectives, an arbitrary number of
behaviors, and capacity (quota) constraints. That paper is the version of record and
should be cited in preference to this one; this preprint is retained for archival
continuity.}

%

\section{Introduction}

Algorithms are increasingly used to translate rich data about individual behavior into consequential decisions affecting peoples' lives. In the process, the prospect of future classification may lead people to change their present behavior in an effort to obtain a better classification outcome.  The prospect of a good credit score, for example, may lead someone to undertake activities that make them more credit worthy. The possibility of an audit may reduce someone's incentives to cheat. Classification algorithms are often designed with these kinds of behavioral goals in mind.

Recent work in machine learning has focused on  \textit{performativity}, or situations in which an algorithm affects the data distribution, and in which  the optimal algorithm depends on this distribution. The notions of \textit{strategic classification} \citep{HardtEtAl16} and \textit{performative prediction} \citep{PerdomoEtAl20} each consider how to classify data that are responsive to an algorithm itself, focusing on the conditions under which it is possible to design an algorithm that properly accounts for performativity.  This literature has largely focused on individuals' efforts to manipulate their data, with individual behavior assumed to be exogenous (a setting termed \textit{data performativity}). A smaller literature considers \textit{outcome performativity} \citep{KimPerdomo23}, in which an individual's true behavioral type may also respond to the algorithm. When outcomes are performative, behavior is endogenous to the algorithm.

This latter setting of outcome performativity is the setting I am concerned with.  In particular, I  assume that the designer of an algorithm knows the data-generating process describing how an individual will respond to algorithmic classification. Anticipating this individual response, an algorithm commits to a classification strategy that will map a signal of the individual's behavior into a classification outcome for the individual.  My question is what an optimal classifier looks like in this context of outcome performativity.

In this setting I  show that the optimal classifier is either a \textit{threshold} or a \textit{negative threshold} rule.  A threshold rule offers the ``good" classification to those individuals whose outcome likelihoods are greater than some cutpoint, $\tau$.  Threshold rules are well-known in the literature on optimal classification and strategic classification \citep{MilliEtAl19,CoateLoury93}, and follow from well-known decision-theoretic results \citep{brown1976complete}.  Negative threshold rules are, to my knowledge, less known. A negative threshold rule offers the ``good" classification outcome to individuals whose outcome likelihoods are \textit{less than} some cutpoint, $\tau$. While seemingly pathological, I show that a negative threshold rule can be the most accurate classifier when outcomes are performative.

These results generalize several recent papers on the topic of classification with endogenous behavior, or outcome performativity. \cite{JungEtAl20} consider the setting of a classification algorithm that is designed to maximize behavioral compliance, showing that the optimal classifier is a (positive) threshold rule with a cutpoint set at the unique crossing point of two behavior-dependent signal distributions. In contrast, my main theorem concerns an accuracy-motivated algorithm, though I provide a corollary extending my result to more general algorithm objectives, and those objectives encompass behavioral compliance-maximization. \cite{PennPatty23,PennPatty24AlgEndog} consider a setting similar to the one considered here but with binary data for the individual (the algorithm simply observes a signal of $0$ or a $1$ for the individual).  In contrast, here I allow the signal space to be the real line. While I model these signals as real numbers arising from two behavior-dependent distributions,  I show that we can equivalently model the signal as representing an outcome likelihood for the individual.  In this sense, the algorithm translates any outcome likelihood in $(0, 1)$ into a classification decision for the individual. 

A foundational aspect of ethical classification is that a classifier should respond in a positively monotonic way to the information provided to it \citep{HardtPriceSrebro16}. In other words, and all else equal,  an individual more likely to be a type ``1" should be more likely to be classified ``1."  Negative threshold rules violate this property, assigning the more desirable classification outcome to the individuals least likely to deserve it.  Such rules can maximize classification accuracy by changing the base rates of behavior in a population. My aim is to pin down the mechanism by which this happens.

\ \\ \textbf{Contributions}

\begin{enumerate}
\item I show that for a general setting of outcome performativity, the most accurate classifier is either a threshold rule or a negative threshold rule. 
\item I provide an example of optimal accuracy being obtained with a negative threshold rule generating classification outcomes that are negatively correlated with signal information.
\item I show that similar results obtain for a more general objective function allowing the algorithm to differentially weigh true positive, true negative, false positive, and false negative classification.

\end{enumerate}

\subsection*{The model and main result}

Consider two players: an individual $i$, and an algorithm, $D$. The individual can take one of two possible actions, $\beta_i\in\{0, 1\}$. We term $\beta_i=1$ as \textit{compliance} and $\beta_i=0$ as \textit{noncompliance}. If choosing $\beta_i=1$ the individual pays cost $\gamma_i$. $\gamma_i$ is private information to the individual, and is drawn from a continuous CDF $H:\mathbf{R}\rightarrow [0, 1]$.

After choosing action $\beta_i$, $D$ observes a signal $x\in\mathbf{R}$ that is drawn from an action-conditional distribution $f_{\beta_i}$. Specifically, let $f_1(x)$ and $f_0(x)$ be two probability density functions that are continuous over the real numbers, with full support. I assume that $f_1(x)$ satisfies the strict monotone likelihood ratio property with respect to $f_0(x)$. Letting $\pi=\Pr[\beta_i=1]$, signal $x$ yields outcome likelihood: $$P(\beta_i=1|x)={\pi{f_1(x)}\over{\pi f_1(x)+(1-\pi)f_0(x)}},$$ and because this likelihood is continuous and strictly increasing, observation of $x$ is equivalent to observation of the outcome likelihood. Note that the assumptions that $H$ is continuous and that  $f_1$ and $f_0$ are continuous with full support are stronger than necessary, but simplify the analysis by allowing us to disregard special cases.


 Finally, upon observing $x$ the algorithm makes a binary decision for $i$,  $d_i\in\{0, 1\}$. $D$'s strategy $\delta(x)$ maps each observed signal into a probability that $i$ is classified as a $1$, or: $$\delta(x)=\Pr [d_i=1|x].$$ I will refer to $\delta(x)$ as a (binary) \textit{classification algorithm}, and assume throughout that $\delta(x)$ is Lebesgue-integrable.

This is a Stackelberg game, as the algorithm commits to a classification strategy prior to the individual's choice of behavior. To summarize, I consider the following timing:

\begin{enumerate}
\item $i$ privately observes behavioral cost $\gamma_i$, drawn from $H$.
\item $D$ commits to classification algorithm $\delta(x)$ with knowledge of cost distribution $H$ and signal distributions $f_0, f_1$.
\item $i$ takes action $\beta_i$ with knowledge of $\delta(x)$ and signal distributions $f_0, f_1$.
\item Signals are generated according to $f_{\beta_i}$ and classified according to $\delta(x)$.
\item Payoffs are received (to be described).
\end{enumerate}

\ \\ \textbf{Payoff to individual}
\ \\If classified as a $1$, $i$ receives a reward $r_1\geq 0$. If classified as a  $0$, $i$ pays a penalty $r_0\leq 0$.  I let $r=r_1-r_0$ be the net benefit to $i$ of being classified as a $1$ versus a $0$, and I assume that $r>0$. Consequently, $i$ chooses $\beta_i=1$ at cost $\gamma_i$ if and only if:

\[ \begin{split} r_1\int_{\mathbf{R}}\delta(x)f_1(x)dx+&r_0 \int_{\mathbf{R}}(1-\delta(x))f_1(x)dx-\gamma_i\geq \\  r_1\int_{\mathbf{R}}\delta(x)& f_0(x)dx+r_0 \int_{\mathbf{R}}(1-\delta(x))f_0(x)dx, \end{split}\]

\ \\which reduces to: 

\begin{equation}\label{behavior}\beta_i=\left\{\begin{array}{rl}1&\text{if}\,\,\, r\int_{\mathbf{R}}(f_1(x)-f_0(x))\delta(x) dx\geq \gamma_i,\\
0&\text{otherwise.}\end{array}\right.\end{equation} Let \begin{equation}\Delta_\delta=\int_{\mathbf{R}}(f_1(x)-f_0(x))\delta(x) dx,\end{equation}  with $\Delta_\delta\in [-1, 1]$ being the difference in probabilities that $i$ is classified as a $1$ if choosing $\beta_i=1$ versus $\beta_i=0$ under algorithm $\delta(x)$. I define $H(r\cdot \Delta_\delta)$ as the \textit{prevalence} induced by classification algorithm $\delta(x)$. It is the ex ante probability that $i$ chooses action $\beta_i=1$ if facing the future prospect of classification according to $\delta(x)$.

\ \\ \textbf{Payoff to algorithm}
\ \\Our algorithm is assumed to be accuracy-maximizing, and so $D$ chooses $\delta$ to maximize: $$\int_{\mathbf{R}}H(r\cdot \Delta_\delta)f_1(x) \delta(x)+(1-H(r\cdot \Delta_\delta))f_0(x)(1-\delta(x))dx.$$

\ \\ Given any algorithm $\delta(x)$, there is a (not necessarily unique) threshold $\tau\in\mathbf{R}$ satisfying:

\begin{equation}\label{equalDelta}\Delta_\delta=\left\{\begin{array}{l c l}\int\limits_{\tau}^\infty (f_1(x)-f_0(x))dx&\text{ if }&\Delta_\delta>0,\\
\int\limits_{-\infty}^{\tau} (f_1(x)-f_0(x))dx&\text{ if }&\Delta_\delta<0.
\end{array}\right.\end{equation}

\ \\ If $\Delta_\delta>0$ the threshold rules solving Equation \ref{equalDelta} must reward signals above some $\tau$; if $\Delta_\delta<0$ the threshold rules must reward signals \textit{below} some $\tau$.

Let $\tau_C$ solve $f_1(x)=f_0(x)$. $\tau_C$ is uniquely defined by the strict MLRP, and it is immediate that a threshold or negative threshold rule with $\tau=\tau_C$ is the unique rule that respectively maximizes or minimizes $\Delta_\delta$. Consequently, if $\tau_C$ solves Equation \ref{equalDelta} then $\delta(x)$ must be a threshold  or negative threshold rule with $\tau=\tau_C$. I state the following Observation without proof, as it is well-known and follows immediately from the fact that the assumptions we've placed on $f_0$ and $f_1$ imply that $$\int\limits_{\tau}^\infty (f_1(x)-f_0(x))dx$$ is strictly quasiconcave with a peak at $\tau_C$, that $$\int\limits_{-\infty}^{\tau} (f_1(x)-f_0(x))dx$$ is strictly quasiconvex with a trough at $\tau_C$, and that both expressions converge to 0 as $\tau\rightarrow\pm\infty$.

\begin{observation}
If $\tau=\tau_C$ does not solve Equation \ref{equalDelta}, by the strict MLRP there are exactly two thresholds, $\tau_L$ and $\tau_H$ solving Equation \ref{equalDelta}, with $\tau_L<\tau_C<\tau_H$.  
\end{observation}

\ \\ The algorithm's payoff from utilizing a positive or negative threshold rule, respectively, that generates prevalence equal to $H(r\cdot\Delta_\delta)$ is the probability $i$ is correctly classified under each of these rules:

\small
$$H(r\cdot \Delta_\delta) \int_{\tau}^{\infty }f_1(x)dx+(1-H(r\cdot \Delta_\delta))\int_{-\infty}^{\tau}f_0(x)dx \,\,\,\text{ if }\Delta>0,$$

$$H(r\cdot \Delta_\delta)\int_{-\infty}^{\tau} f_1(x)dx+(1-H(r\cdot \Delta_\delta))\int_{\tau}^\infty f_0(x)dx \,\,\,\text{ if }\Delta<0. $$

\normalsize

\ \\Our goal is to show that one of these threshold rules is always weakly more accurate than $\delta(x)$. 

\ \\ \begin{theorem}Threshold or negative threshold rules are optimally accurate for classification with performativity. Specifically:\end{theorem}
\textit{\begin{itemize}
\item Let $i$'s behavior $\beta_i$ be performative (i.e. depend on the prospect of classification according to $\delta(x)$) in the sense of satisfying Equation \ref{behavior}. 
\item Let signals of behavior $x$ be generated according to $f_{\beta_i}$, with  $f_1$ satisfying the strict MLRP with respect to $f_0$, and $f_1, f_0$ continuous with full support.
\end{itemize} For any integrable classification algorithm $\delta(x):\mathbf{R}\rightarrow[0, 1]$, there exists either a threshold rule or a negative threshold rule that is as accurate as $\delta(x)$.}

\ \\ \textit{Proof}: Our proof proceeds in two steps. In \textbf{Step 1} I derive a necessary and sufficient  condition for a threshold rule (and respectively, a negative threshold rule) to be as accurate as classification algorithm $\delta(x)$. In \textbf{Step 2} I show that this condition always holds.

\ \\ \textbf{Step 1}: Fix $\delta(x):\mathbf{R}\rightarrow[0, 1]$ to be any classification algorithm. Let:

$${\delta}_0=\int_{\mathbf{R}}f_0(x)\delta(x)dx\,\,\,\text{ and }\,\,\,{\delta}_1=\int_{\mathbf{R}}f_1(x)\delta(x)dx.$$

\ \\I begin by defining the following functions $R^{\pm}_0(\tau)$ and $R^\pm_1(\tau)$:

\small 
\begin{align*}
R_0^+(\tau) &= \delta_0 - \int_{\tau}^{\infty} f_0(x) \, dx, &
R_1^+(\tau) &= \delta_1 - \int_{\tau}^{\infty} f_1(x) \, dx, \\
R_0^-(\tau) &= \delta_0 - \int_{-\infty}^{\tau} f_0(x) \, dx, &
R_1^-(\tau) &= \delta_1 - \int_{-\infty}^{\tau} f_1(x) \, dx.
\end{align*} \normalsize These functions are ``remainder" terms, with $R_{\beta_i}^+(\tau)$ representing the difference in probability that an individual who has chosen $\beta_i$ is classified as $d_i=1$ under classifier $\delta(x)$ versus under a threshold rule with threshold $\tau$.  $R_{\beta_i}^-(\tau)$ is defined similarly for negative threshold rules.

We'll first consider the case of $\Delta_\delta>0$, letting $H\equiv H(r\cdot\Delta_\delta)$ throughout. For our threshold rule to be as accurate as $\delta(x)$ we need Equation \ref{posT} to be nonnegative:

\begin{equation}\label{posT}\begin{split}  H \int_{\tau}^{\infty }f_1(x)dx+&(1-H)\int_{-\infty}^{\tau}f_0(x)dx \\-\int_{\mathbf{R}}Hf_1(x) \delta(x)&-(1-H)f_0(x)(1-\delta(x))dx.\end{split}\end{equation}

\normalsize
 
 \ \\ We can decompose Equation \ref{posT}  into the following two parts representing the accuracy difference between the threshold and optimal rule, $\delta(x)$:

\begin{equation}\label{posAccuracy}\begin{array}{l}-HR_1^+(\tau)+(1-H) R_0^+(\tau).\end{array}\end{equation}

\ \\By the fact that $\tau\in\{\tau_L, \tau_H\}$ yields identical prevalence as $\delta(x)$ and $\Delta_\delta>0$ we have:
$$\delta_1-\delta_0=\int\limits_{\tau}^\infty (f_1(x)-f_0(x))dx,\text{ or }$$

\begin{equation}\label{equal+}R_1^+(\tau)=R_0^+(\tau)\end{equation}for $\tau \in\{\tau_L, \tau_H\}.$

\ \\Finally, Equations \ref{posAccuracy} and \ref{equal+} show that if $\Delta_\delta>0$ and the following condition holds, the threshold rule is as accurate as $\delta(x)$. 

\begin{condition}\label{posAccurate}
$$\boxed{\begin{array}{rcl}H\leq{1\over 2}&\text{ and }&  R_0^+(\tau_H) \geq 0, \text{ or }\\
H\geq{1\over 2}&\text{ and }&R_1^+(\tau_L) \leq 0.
\end{array}}$$
\end{condition}

\ \\  We'll next consider the case with $\Delta_\delta<0$, again letting $H\equiv H(r\cdot\Delta_\delta)$ throughout.  For our negative threshold rule to be as accurate as $\delta(x)$, we need Equation \ref{negT} to be nonnegative:

\begin{equation}\label{negT} \begin{split}H\int_{-\infty}^{\tau} f_1(x)dx&+(1-H)\int_{\tau}^\infty f_0(x)dx \\ -\int_{\mathbf{R}}&Hf_1(x) \delta(x)-(1-H)f_0(x)(1-\delta(x))dx. \end{split} \end{equation}

\ \\ Again, I separate Equation \ref{negT} into two components, representing the accuracy difference between the negative threshold rule and $\delta(x)$:

\begin{equation}\label{negAccuracy}\begin{array}{l}-HR_1^-(\tau)+(1-H) R_0^-(\tau).\end{array}\end{equation}

\ \\By the fact that $\tau\in\{\tau_L, \tau_H\}$ yields identical prevalence as $\delta(x)$ and $\Delta_\delta<0$ we have:
$$\delta_1-\delta_0=\int_{-\infty}^{\tau}f_1(x)-f_0(x)dx<0,\text{ or }$$
\begin{equation}\label{eqR}R_1^-(\tau)=R_0^-(\tau),\end{equation} for $\tau\in\{\tau_L, \tau_H\}.$

\ \\ Finally, Equations \ref{negAccuracy} and \ref{eqR} show that if $\Delta_\delta<0$ and either of the following hold then the negative threshold rule is as accurate as $\delta(x)$. 

\begin{condition}\label{negAccurate}
$$\boxed{\begin{array}{rcl}H\leq{1\over 2}&\text{ and }&  R_0^-(\tau_H) \leq 0, \text{ or }\\
H\geq{1\over 2}&\text{ and }&R_1^-(\tau_L) \geq 0.
\end{array}}$$
\end{condition}

\vspace{.25in}

\ \\\textbf{Step 2}: We'll now show that  if $\Delta_\delta>0$ then  Condition \ref{posAccurate} holds. If $\Delta_\delta<0$ then  Condition \ref{negAccurate} holds by a symmetric argument.  

\ \\ Suppose that $\tau_C$ is not a solution to Equation \ref{equalDelta} (if it is a solution, we know that $\delta(x)$ must itself be a threshold rule).  Assume without loss of generality that $\Delta_\delta>0$; the $\Delta_\delta<0$ case follows symmetrically. We'll show that it must be the case that $R^+_1(\tau_L)\leq 0$ and $R^+_0(\tau_H)\geq 0$. I start by showing that  $R_0^+(\tau_H) \geq 0$. 

\ \\ Let \( h(x) = f_1(x) - f_0(x) \). Since \( f_1 / f_0 \) is increasing, define:
\[
g(x) = \frac{h(x)}{f_0(x)} = \frac{f_1(x)}{f_0(x)} - 1.
\]
Then \( g(x) \) is strictly increasing, and \( g(\tau_C) = 0 \), with:
\[
g(x) < 0 \text{ for } x < \tau_C, \qquad g(x) > 0 \text{ for } x > \tau_C.
\]  

\ \\ Define the function:
\[
\eta(x) = \delta(x) - \mathbf{1}_{x > \tau_H},
\] and note that $\eta(x)\leq 0$ for $x>\tau_H$, $\eta(x)\geq 0$ for $x<\tau_H$, and $\eta(x)\in[-1, 1]$.  Then:

\[
R_0^+(\tau_H) =\delta_0 - \int_{\tau_H}^\infty f_0(x) \, dx = \int_{\mathbf{R}} f_0(x) \eta(x) \, dx,
\]
and using \( h(x) = f_0(x) g(x) \), we can write:
\small 
\begin{equation}\label{eta0}
\Delta_\delta - \int_{\tau_H}^{\infty} h(x) dx = \int_{\mathbf{R}} h(x) \eta(x) \, dx = \int_{\mathbf{R}}  g(x) \eta(x)f_0(x) \, dx = 0.
\end{equation}
\normalsize 
\ \\ Define:
\[
A = \{ x < \tau_H : \eta(x) > 0 \}, \quad B = \{ x > \tau_H : \eta(x) < 0 \}.
\]
By decomposing Equation \ref{eta0} into two parts we have:
\[\begin{split}
\int_{\mathbf{R}} g(x) \eta(x) f_0(x) dx = &\\ \int_A g(x) \eta(x) f_0(x) dx + &\int_B g(x) \eta(x) f_0(x) dx = 0.
\end{split}
\]

\ \\ Because $g$ is strictly increasing we have that for all \( x \in A \),  \( g(x) < g(\tau_H) \), and for all \( x \in B \), \( g(x) > g(\tau_H) \).

\ \\We can write:
\[\begin{split}
\int_A g(x) \eta(x) f_0(x) dx = &\\  \int_A (g(x) - g(\tau_H)) &\eta(x) f_0(x) dx + g(\tau_H) \int_A \eta(x) f_0(x) dx,
\end{split}
\]
\[\begin{split}
\int_B g(x) \eta(x) f_0(x) dx =&\\ \int_B (g(x) - g(\tau_H)) &\eta(x) f_0(x) dx + g(\tau_H) \int_B \eta(x) f_0(x) dx.
\end{split}
\] 

\ \\ Note that:
$$g(x) - g(\tau_H) < 0 \,\,\text{ and }\,\,\eta(x) > 0  \,\,\text{ on }  A$$ implies  $$ \int_A (g(x) - g(\tau_H)) \eta(x) f_0(x) dx < 0,$$ and 
$$g(x) - g(\tau_H) > 0 \,\,\text{ and }\,\,\eta(x) < 0 \,\,\text{ on } B $$ implies $$     \int_B (g(x) - g(\tau_H)) \eta(x) f_0(x) dx < 0.$$

\ \\ Therefore,
\begin{equation} \label{inequalities} \begin{array}{c}
\int_A g(x) \eta(x) f_0(x) dx < g(\tau_H) \int_A \eta(x) f_0(x) dx,\\
\\
\int_B g(x) \eta(x) f_0(x) dx < g(\tau_H) \int_B \eta(x) f_0(x) dx.
\end{array}
\end{equation}

\ \\ Adding the left and right sides of the inequalities in Equation \ref{inequalities} we get that:

\[\begin{split}
0 = \int_A g(x) \eta(x) f_0(x) dx + \int_B g(x) \eta(x) f_0(x) dx&\\
< g(\tau_H) \big( \int_A \eta(x) f_0(x) dx + \int_B \eta(x)  f_0(x) &dx \big).\end{split}
\]
\normalsize
\ \\ This, along with the fact that $g(\tau_H)>0$, implies:
\[
R_0^+(\tau_H) =\int_{\mathbf{R}} \eta(x) f_0(x) dx \geq 0,
\] which is what we sought to show.

\ \\We prove the remaining cases using identical arguments, defining $\eta(x)$ differently for each case. To show \( R_1^+(\tau_L) \le 0 \) let \( \eta(x) = \delta(x) - \mathbf{1}_{x > \tau_L} \).   To show  \( R_1^-(\tau_L) \ge 0 \),  let \(\eta(x) = \delta(x) - \mathbf{1}_{x < \tau_L} \).  And to show  \( R_0^-(\tau_H) \le 0 \), let \( \eta(x) = \delta(x) - \mathbf{1}_{x <\tau_H} \).

\ \\ Finally, if $\Delta_\delta=0$ then $$\int_{\mathbf{R}}f_1(x)\delta(x)dx=\int_{\mathbf{R}}f_0(x)\delta(x)dx.$$ Therefore the accuracy of $\delta(x)$ is: $$\int_R \left(H(0)\delta(x)+(1-H(0))(1-\delta(x)) \right) f_1(x)dx,$$ and accuracy is maximized by setting: 
$$\delta(x)= \left\{ \begin{array}{rcl} 1 &\text{ if }&H(0)\geq {1\over 2}\\
0&\text{if}&H(0)\leq {1\over 2},\end{array}\right.$$ which is a threshold rule with $\tau\in\{-\infty, \infty\}$.

\ \\ It follows that for any strategy $\delta(x)$,  if $\Delta_\delta<0$ then  Condition \ref{negAccurate} holds and if $\Delta_\delta>0$ then  Condition \ref{posAccurate} holds.  If $\Delta_\delta=0$ then $\delta(x)$ is a constant function with $\delta(x)\in\{0, 1\}$, $\forall x$. Consequently, there exists a threshold or negative threshold rule that is as accurate as $\delta(x)$. $\Box$

\subsection*{Example of a most-accurate negative threshold rule}
In this section I provide an example of an environment in which a negative threshold rule is more accurate than a positive threshold rule due to the performativity of the classifier.

\ \\Suppose that the individual's cost is distributed $\gamma_i\sim \mathcal{N}[{3\over 4}, 1]$, that the stakes to classification $r=r_1-r_0=5$, and that the signal distribution $f_{\beta_i}$ is  $\mathcal{N}[\beta_i, 1]$, for $\beta_i\in\{0, 1\}$.  

The {accuracy-maximizing  positive threshold rule} sets $\tau\approx-0.1$. Letting $H$ be the CDF of the individual's cost distribution, the probability that $i$ chooses $\beta_i=1$ at this classifier is $$H\left(5\cdot \int_{-0.1}^\infty f_1(x)-f_0(x)dx \right)\approx H(1.625) \approx 0.81.$$ The accuracy of this positive threshold classifier is: $$0.81 \int_{-0.1}^\infty f_1(x) dx+0.19\int_{-\infty}^{-0.1} f_0(x) dx\approx 0.787.$$

The {accuracy-maximizing negative threshold rule} sets $\tau\approx-1.4$. The probability that $i$ chooses $\beta_i=1$ at this classifier is $$H\left(5\cdot \int_{-\infty}^{-1.4} f_1(x)-f_0(x)dx \right)\approx H(-0.36) \approx 0.13.$$  The accuracy of this negative threshold classifier is: $$0.13 \int_{-\infty}^{-1.4} f_1(x) dx+0.87\int_{-1.4}^{\infty} f_0(x) dx\approx 0.801$$

\ \\It follows that the negative threshold rule yields a more accurate classification outcome than the positive threshold rule. This is due to the outcome performativity of the classifier; the negative threshold rule induces greater behavioral non-compliance by the individual (an 87\% probability that $\beta_i=0$) than the greater behavioral compliance induced by the positive threshold (an 81\% probability that $\beta_i=1$). This downward shift in the individual's base rate facilitates more accurate classification. By our Theorem, the negative threshold rule is the \textit{most accurate} classifier for this example.

\subsection*{More general algorithms}

So far I've assumed that the algorithm seeks to maximize accuracy. However, the result that optimal classifiers are threshold or negative threshold rules can be extended to cover a richer set of classifier objectives. Now suppose that the algorithm chooses $\delta(x)$ to maximize the following more general objective function, letting the terms $A_1, B_1, A_0, B_0\in\mathbf{R}$.

\begin{equation}\begin{split}
\int_{\mathbf{R}}\left(\delta(x)A_1+(1-\delta(x))A_0 \right) f_1(x)dx&\\ +\int_{\mathbf{R}}\left((1-\delta(x))B_1+\delta(x)B_0 \right)&f_0(x) dx.\end{split}
\end{equation}

\begin{table}[hbtp]
    \centering
    \begin{tabular}{|c||c|c|} \hline
&\multicolumn{2}{c|}{Decision} \\ \hline
Behavior&$d_i=1$&$d_i=0$ \\ \hline \hline

\multirow{2}{*}{$\beta_i=1$}&$A_1$&$A_0$ \\ 
& (True Positive)&(False Negative) \\ \hline
\multirow{2}{*}{$\beta_i=0$}&$B_0$&$B_1$\\ 
& (False Positive) & (True Negative) \\ \hline
 \end{tabular}

\end{table} \noindent Consequently, the algorithm receives a payoff that differentially weights the probability that $i$ falls into any of the four cells of the confusion matrix. Our accuracy-maximizing classifier set $A_1=B_1=1$ and $A_0=B_0=0$. A compliance-maximizing classifier would set $A_1=A_0=1$ and $B_1=B_0=0$. This more general framework allows the algorithm to differentially weigh true positives, true negatives, false positives, and false negatives. Note that the optimization problem of the algorithm is unique up to any positive affine transformation of the values $\{A_1, A_0, B_1, B_0\}$.

Consider the following restriction on the objectives of the algorithm, as in \cite{PennPatty23,PennPatty24AlgEndog}. These restrictions require that, conditional on behavior $\beta_i$, the algorithm weakly prefers either more accurate classification or less accurate classification.

\begin{definition}Algorithm $D$ is \textit{accuracy-aligned} if $[A_1\geq A_0$ and $B_1\geq B_0]$.  $D$ is \textit{accuracy-misaligned} if $[A_1\leq A_0$ and $B_1\leq B_0]$. 
\end{definition} Note that accuracy-maximization and compliance-maximization are both instances of accuracy-alignment. We're now ready to state a corollary to our theorem.

\begin{corollary} Let $D$'s objectives  be either accuracy-aligned or accuracy-misaligned. Then a threshold or negative threshold rule is optimal for classification with performativity. 
\end{corollary}

\noindent  \textit{Proof}: If $A_1=A_0$ and $B_1=B_0$, then $D$ is compliance-maximizing (maximizing $H(r\cdot \Delta_\delta)$) or compliance-minimizing (minimizing $H(r\cdot\Delta_\delta)$).  Consequently, the optimal classifier is a threshold or negative threshold setting $\tau=\tau_C$. 

We'll now assume that either $A_1\not=A_0$ or $B_1\not=B_0$ or both. Fix any $\delta(x)$ with $\Delta_\delta>0$, again letting $H\equiv H(r\cdot\Delta_\delta)$. For our threshold rule to yield as high a payoff as $\delta(x)$ we need Equation \ref{posGen} to be nonnegative:

\begin{equation}\begin{split}\label{posGen}  H \big(A_1 \int_{\tau}^{\infty }f_1(x)dx  +A_0 \int_{-\infty}^{\tau}f_1(x)dx  \big)  & \\ + (1-H)\big(B_1\int_{-\infty}^{\tau}f_0(x)dx+ B_0 \int_{\tau}^\infty & f_0(x)dx\big)\\
-H\int_{\mathbf{R}}\big(A_1\delta(x)+A_0(1-\delta(x))\big) &f_1(x) dx\\
-(1-H)\int_{\mathbf{R}}\big(B_1(1-\delta(x))+&B_0\delta(x) \big)f_0(x)dx.\end{split}
\end{equation}

\ \\ Re-expressing Equation \ref{posGen}, we get that the positive threshold rule yields as high a payoff as $\delta(x)$ when: \begin{equation}\label{posInter}(1-H)(B_1-B_0)R^+_0(\tau)-H(A_1-A_0)R_1^+(\tau)\geq 0.\end{equation} Equation \ref{posInter} yields the following Condition \ref{genCond}, an analog of Condition \ref{posAccurate}. If Condition \ref{genCond} is satisfied, a positive threshold rule yields as high payoff as $\delta(x)$.

\begin{condition}\label{genCond}
\scriptsize 
\[\boxed{\begin{aligned}H\leq{{B_1-B_0}\over{A_1-A_0+B_1-B_0}} &\text{ \& }   R_0^+(\tau_H) \geq 0  \text{ \& }[A_1\geq A_0 \& B_1\geq B_0], \text{ or }\\
H\geq{{B_1-B_0}\over{A_1-A_0+B_1-B_0}} &\text{ \& }R_1^+(\tau_L) \leq 0\text{ \& }[A_1\geq A_0 \& B_1\geq B_0], \text{ or }\\
H\geq{{B_1-B_0}\over{A_1-A_0+B_1-B_0}} &\text{ \& }  R_0^+(\tau_H) \geq 0 \text{ \& }[A_1\leq A_0 \& B_1\leq B_0], \text{ or }\\
H\leq{{B_1-B_0}\over{A_1-A_0+B_1-B_0}} & \text{ \& }R_1^+(\tau_L) \leq 0\text{ \& }[A_1\leq A_0 \& B_1\leq B_0].
\end{aligned} }\]
\end{condition} \noindent Finally, Step 2 of our Theorem proves that Condition \ref{genCond} is always satisfied. The case of $\Delta_\delta<0$ is proved similarly. $\Box$
\normalsize

\section{Conclusion}

Let \textit{score} $s_i(x)=P(\beta_i=1|x)$ be the the probability that person $i$ has chosen behavior $\beta_i=1$ conditional on information $x$. A classifier satisfies  \textit{score monotonicity} if $s_i(x)\geq s_i(y)$ implies $\Pr(d_i=1|x)\geq \Pr(d_i=1|y)$. In  words, if information $x$ leads us to believe that person $i$ was more likely to have complied than information $y$ would have led us to believe, then person $i$ should be at least as likely to obtain a positive classification outcome under information $x$ than information $y$. I've shown that \textit{optimal classification can violate score monotonicity}. When this occurs, it is because an optimal classifier can successfully drive down behavioral compliance in order to improve its own accuracy.

These results have implications for the ethical use of algorithmic classification---particularly in the absence of auditing mechanisms. They also point to a fundamental conflict between accuracy and intuitive fairness, trust, and interpretability. Intuitively, people with stronger evidence for belonging to a positive class shouldn't be less likely to be classified positively than those with weaker evidence. Stakeholders should expect that ``better" candidates aren't unfairly passed over in favor of ``worse" ones.  And legally, violations of score monotonicity can look like capricious or arbitrary discrimination.  

Perhaps more distressingly, because an optimal classifier anticipates behavioral responses to its own classification strategy, it attempts to manipulate behavior in order to maximize its own objective function. Pursuit of a seemingly neutral goal such as accuracy can lead a classifier to incentivize individuals to make negative life choices. While threshold-based classification rules are well-known, widely studied, and intuitively fair, my goal has been to show that negative thresholds can also be optimal in a simple setting of classification with outcome performativity, in which behavior responds to the algorithm. That such rules are not well-known, not widely studied, and intuitively unfair is notable.

\bibliography{john}

\end{document}